\documentclass[conference,10pt]{IEEEtran}

\usepackage{times}

\usepackage{amsthm}
\usepackage{amsmath}    
\usepackage{amssymb}    
\usepackage{mathrsfs}   
\usepackage{epsfig}     
\usepackage{graphicx}

\usepackage{url}

\usepackage{comment}
\usepackage{booktabs}

\usepackage{subfigure}

\usepackage{amsfonts}
\usepackage{dsfont}

\usepackage[numbers,sort&compress]{natbib}

\usepackage{enumerate}

\usepackage[hmargin=1.3cm,vmargin=1.3cm]{geometry}

\newcommand{\vep}{\varepsilon}

\newcommand{\B}{{\mathcal B}}

\renewcommand{\S}{{\mathcal S}}

\newcommand{\T}{{\mathcal T}}

\newcommand{\Rs}{{\mathcal R}}

\newcommand{\s}{{\bf s}}
\renewcommand{\r}{{\bf r}}
\newcommand{\x}{{\bf x}}
\newcommand{\y}{{\bf y}}

\newcommand{\lcrit}{\ell_{\rm crit}}

\newcommand{\lacrit}{\tilde \ell_{\rm crit}}

\newcommand{\Rsd}{\mathcal{R}_{L,D}}

\newcommand{\eref}[1]{(\ref{#1})}

\newcommand{\linter}{\ell_{\rm inter}}
\newcommand{\lth}{\ell_{\rm th}}

\newtheorem{theorem}{Theorem} 
\newtheorem{lemma}{Lemma} 
\newtheorem{claim}{Claim} 
\newtheorem{cor}{Corollary} 

\newtheorem{remark}{Remark}[section] 
\newtheorem{question}{Question} 

\newcommand{\aln}[1]{\begin{align*}#1\end{align*}}

\newcommand{\al}[1]{\begin{align}#1\end{align}}

\begin{document}
%
\title{Do Read Errors Matter for Genome Assembly?}

\author{  
   \IEEEauthorblockN{Ilan Shomorony}  
   \IEEEauthorblockA{UC Berkeley \\ ilan.shomorony@berkeley.edu   \vspace{-4mm}
   }
  \and
   \IEEEauthorblockN{Thomas Courtade}
   \IEEEauthorblockA{UC Berkeley \\ courtade@berkeley.edu  \vspace{-4mm}
   }
  \and
   \IEEEauthorblockN{David Tse}
   \IEEEauthorblockA{Stanford University \\ dntse@stanford.edu}  \vspace{-4mm}
 }

\maketitle

\begin{abstract}
While most current high-throughput DNA sequencing technologies generate short reads with low error rates, emerging sequencing technologies generate long reads with high error rates. 
A basic question of interest is the tradeoff between read length and error rate in terms of the information needed for the perfect assembly of the genome. 
Using an adversarial erasure error model, we make progress on this problem by establishing a critical read length, as a function of the genome and the error rate, above which perfect assembly is guaranteed. 
For several real genomes, including those from the GAGE dataset,  we verify that this critical read length is not significantly greater than the read length required for perfect assembly from reads without errors. 
\end{abstract}

%

\section{Introduction}
\label{sec:intro}

Current DNA sequencing technologies are based on a two-step process.
First, tens or hundreds of millions of fragments from random locations on the DNA sequence are read via \emph{shotgun sequencing}.
Second, these fragments, called reads, are merged to each other based on regions of overlap, using an \emph{assembly algorithm}. 

Roughly speaking, different shotgun sequencing platforms can be distinguished from the point of view of three main metrics: the \emph{read length}, the 
\emph{read error rate}, 
and the \emph{read throughput}.
In the last decade, the so-called next-generation sequencing platforms 
have attained considerable success at employing heavy
parallelization in order to achieve \emph{high-throughput} shotgun sequencing. 
This allowed a significant reduction in the cost and time of sequencing, causing an explosion in the number of new sequencing projects and the generation of massive amounts of sequencing data.


In order to guarantee low error rates, most of these next-generation technologies are restricted to \emph{short read lengths}, shifting some of the burden of sequencing to the assembly step. 
In practice, this results in very fragmented assemblies, with large gaps and little linking information between fragments \cite{MindTheGaps}.
On the other hand, recent technologies that generate longer reads suffer from lower throughput and much higher error rates\footnote{One example of a short-read-length technology is Illumina, with reads of length $\sim 200$ base pairs and error rates of about $1\%$.
In contrast, PacBio reads can be several thousand base pairs long, with error rates of about $10$-$15\%$.}.

Given this technology trend, the natural questions to ask are: what is the impact of read errors on the performance of assemblers? 
Is the negative impact of read errors 
more than offset 
by the increase in read lengths in long-read technologies? 
It is well known that read errors have a significant impact on assembly algorithms. 
For example, in DeBruijn graph based algorithms, read errors create extraneous nodes and edges in the assembly graph, 
which results in added complexity.
However, these observations  pertain to {\em specific} algorithms. 
A more fundamental question can be asked from an {\em  information-theoretic} point of view: 
given a read length, an error rate and a coverage depth (number of reads per base), is there enough {\em information} in the read data to uniquely reconstruct the genome?
Do errors significantly increase the read length and/or coverage depth requirements? 
An answer to these basic feasibility questions can provide an algorithm-independent framework for evaluating different sequencing technologies. 
It would also settle some speculations  in the assembly community on whether read errors have a significant impact in long-read technologies (see for example \cite{MyersTweet}).

\begin{figure}[b] 
\vspace{-3mm}
	\center
       \hspace{0mm}
                \subfigure[\emph{S. aureus}]{
                \hspace{-6mm}
       \includegraphics[width=0.49\linewidth]{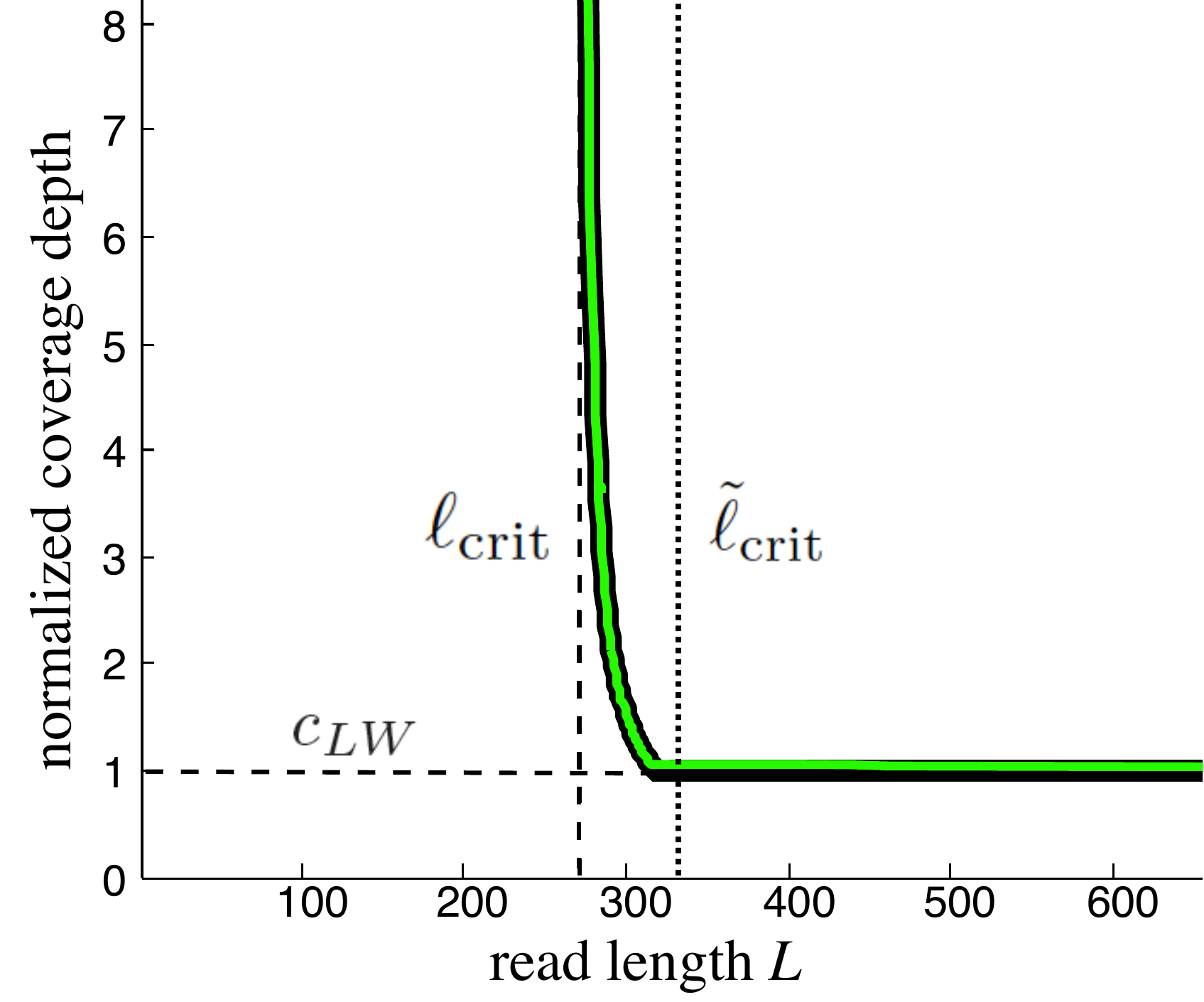} \label{tworeadsfig}} 
       \hspace{-5mm}
                \subfigure[\emph{R. sphaeroides}]{
       \includegraphics[width=0.49\linewidth]{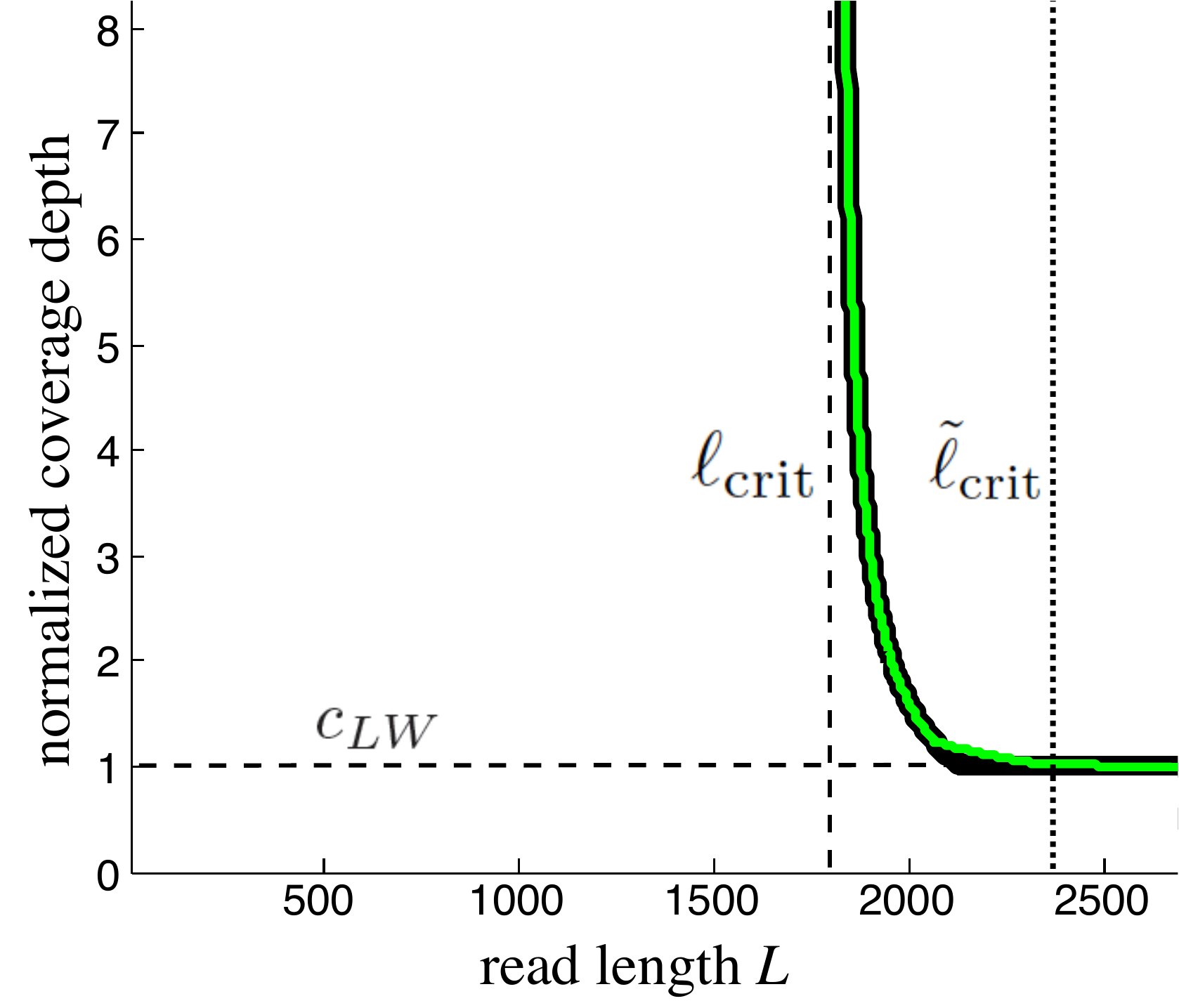} \label{outedgesfig}}
       \caption{The thick black curve is a feasibility lower bound for any algorithm, and the green line represents the performance of the Multibridging algorithm \cite{BBT}.
       \label{fig:feasibility} }
\end{figure}

\begin{figure}[t] 
	\center
       \includegraphics[width=0.98\linewidth]{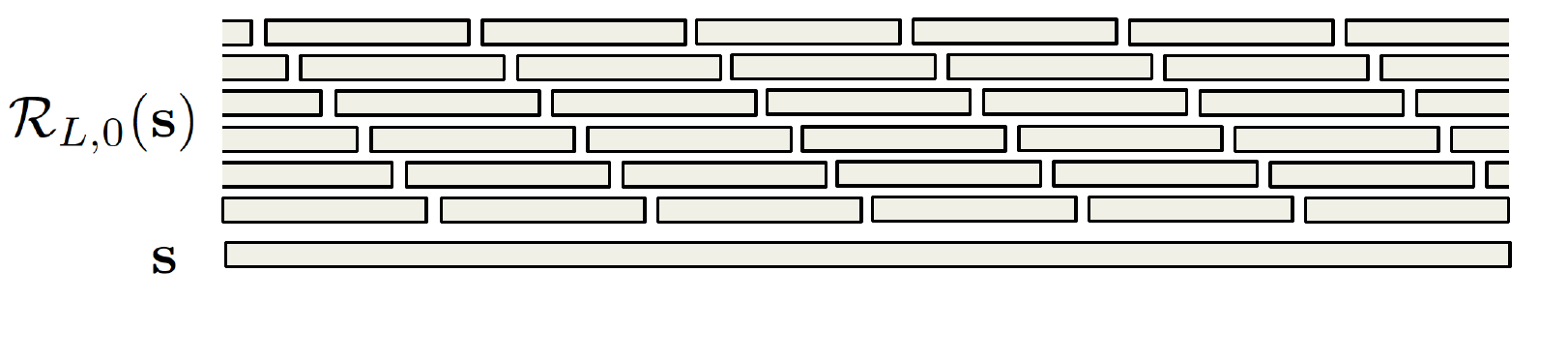} \vspace{-4mm}
        \caption{The sequence $\s$ and its $L$-spectrum, $\Rs_{L,0}(\s)$.
        \label{densereadsfig}}
        \vspace{-4mm}
\end{figure}

Such a framework was initiated in \cite{BBT} for {\em error-free} reads: a feasibility curve relating the read length and coverage depth needed to perfectly assemble a genome was characterized in terms of the repeat complexity of the genome (see examples in Fig.~\ref{fig:feasibility}).
Evaluating this curve on several genomes revealed an interesting threshold phenomenon: 
if the read length is below a certain critical value $\lcrit$, reconstruction is impossible;  a read length slightly above $\lcrit$ and a coverage depth close to the Lander-Waterman depth $c_{LW}$ (i.e., just enough reads to cover the whole sequence) is sufficient.
The critical read length  $\lcrit$ is given by the length of the longest \emph{interleaved repeat} in the genome, and coincides with the minimum read length $L$ needed to uniquely reconstruct the genome given its $L$-\emph{spectrum}, i.e. the set of reads with one length-$L$ read starting at each position of the sequence, illustrated in Fig.~\ref{densereadsfig}. 
This minimum read length also appeared in earlier works by Ukkonen and Pevzner \cite{Ukkonen,PevznerDNA} for reconstruction via \emph{sequencing by hybridization}. 

Given this framework, the impact of read errors can be studied by asking how much the critical read length $\lcrit$ increases when there are errors. 
In this paper, we investigate this tradeoff
for a specific error model:
1) the errors are erasures; 
2) the erasures occur at a rate no larger than $D/L$ for each read and for each base in the sequence, but are otherwise arbitrary. 
Our main result is the characterization of a critical read length $\lacrit$ 
above which perfect assembly is always possible.
While in the noiseless case $\lcrit$ is a function of the sequence repeat structure, $\lacrit$ depends more generally on the error rate and on the \emph{approximate repeats} in the sequence.
More concretely, for a sequence $\s$,
\aln{ 
\lacrit(\s,D) = \min_{k \geq \lcrit(\s)} k + D \cdot M_{\s}(D,k+1), 
}
where $M_{\s}(D,\ell)$ is the maximum number of $D$-approximate length-$\ell$ repeats in $\s$.
Moreover, reminiscent of classical coding theory results, 
we show that the same read length $\lacrit$ is sufficient for assembly if instead of erasures we consider substitution errors at half of the rate.
In order to characterize $\lacrit$, we derive a new result about the error correction capability of the $L$-spectrum.
More precisely, we show that given a noisy version of the $L$-spectrum of a sequence, it is possible to obtain the noiseless $(k+1)$-spectrum of the same sequence, for any $k$ such that $L > k + D \cdot M_{\s}(D,k+1)$.
When $L > \lacrit$, we can obtain the noiseless $(k+1)$ spectrum for some $k > \lcrit$,
and the noiseless result from \cite{BBT} implies that perfect assembly is possible.

By evaluating $\lacrit$ on several real genomes, including those in the GAGE dataset \cite{GAGE}, we verify that $\lacrit$ is not significantly larger than $\lcrit$.
In fact, in most cases, $\lacrit \approx \lcrit + 3D$.
Hence, if the read length $L$ is chosen above the noiseless requirement $\lcrit$, 
perfect assembly is robust to errors up to a threshold 
(roughly $\frac13(L-\lcrit)$ erasures per read).

The impact of read errors on the information theoretic limits of genome assembly has also been studied in the setting of an i.i.d.~genome model and asymptotically long genome length \cite{MotahariNoisy}, building on an earlier work on error-free reads in the same setting \cite{MotahariDNA}.  
The results are surprising: as long as the error rate is below a threshold (which can be as high as $19 \%$ for substitution errors), noisy reads are as good as noiseless reads; i.e., the requirements for assembly in terms of read length and coverage depth are the same in both cases. 
While this result seems stronger than the result in the present paper, it is proved under the idealistic and unrealistic settings of i.i.d.~genome statistics and i.i.d.~errors. 
The present result, on the other hand, is more robust as it applies to arbitrary genome repeat statistics and error statistics.

\section{Problem Setting}
\label{probsec}

In the DNA assembly problem, the goal is to reconstruct a sequence $\s = (s[1],...,s[G])$ 
of length $G$ with symbols from the alphabet $\Sigma = \{a,c,g,t\}$.
In order to simplify the exposition, we assume a \emph{circular} DNA model;
thus, $\{s[i]\}_{i=1}^\infty$ is a periodic sequence with (minimum) period $G$.
Our results hold in the non-circular case as well under minor modifications.
We will let $\s_i^\ell$ be the substring of length $\ell$ starting at $s[i]$; i.e., $\s_i^\ell = (s[i],s[i+1],...,s[i+\ell-1])$.

The sequencer provides a multiset of $N$ \emph{reads} $\Rs = \{\r_1,...,\r_N\}$ from $\s$, each of length $L$.
In the noiseless case, each read is a length-$L$ substring of $\s$ with an unknown starting location.
Our focus, however, will be on noisy read models, 
where each read may be corrupted by noise.
The goal is to design an \emph{assembler}, which takes the set of reads $\Rs$ and attempts to reconstruct the sequence $\s$.


\subsection{The $L$-Spectrum Read Model}

We will consider a ``dense-read'' model, in which all the reads in the $L$-spectrum of $\s$ are provided.
More precisely, $\Rs$ will have exactly $G$ reads, one from each possible starting position; i.e., $\Rs = \{\r_1,...,\r_G\}$, where $\r_i = \s_i^L$ for $i=1,...,G$.
We will refer to the error-free $L$-spectrum of $\s$ by $\Rs_{L,0}(\s)$.
Notice that the starting position $i$ for each read $\r_i$ is unknown to the assembler.

While such a read model was originally proposed in the context of \emph{sequencing by hybridization} \cite{Ukkonen,PevznerDNA,seqhyb}, 
our motivation for using it comes from next-generation sequencing technologies, where the high read throughput can provide large coverage depths at low costs, and a dense read regime is not unrealistic.
This way, we can bypass the question of the necessary coverage depth for assembly, and
instead focus on the interplay between read length and error rate in the context of assembly feasibility.
Moreover, as shown in \cite{BBT} for noiseless reads, the dense-read model provides valuable insights towards understanding the information-theoretic limits of reconstruction in the more general shotgun read model.

In the $L$-spectrum read model, since we have exactly $G$ reads, an assembly of the reads $\Rs_{L,0}(\s) = \{\r_1,...,\r_G\}$ can be thought of as a permutation $\sigma$ of the entries of $(1,...,G)$.
We assume without loss of generality that the identity permutation $\sigma_0 = (1,...,G)$ yields a correct assembly of $\s$.
Notice, however, that the index $i$ of each read $\r_i$ is unknown to the assembler.
Notice also that in general, there may be multiple correct assemblies for a sequence $\s$ if  $\r_i = \r_j$ for some $i \ne j$.

\subsection{Adversarial Erasure Model} \label{advsec}

As in the classical coding theory literature, we will study the problem of DNA assembly with noisy reads from the perspective of an \emph{adversarial} noise model.
Given that actual sequencing noise profiles are complex (non-i.i.d., asymmetric across bases) and technology-dependent, this approach avoids the need for a probabilistic noise model by instead focusing on a worst-case scenario.
Moreover, under this model we can hope to obtain deterministic and non-asymptotic conditions for perfect assembly, which can be more easily analyzed in terms of real genome data.

Motivated by the fact that sequencing technologies usually provide a quality score for each base that is read (which could be thresholded into ``good'' and ''bad'' bases), and in order to simplify the problem, we will consider an \emph{erasure} model.
The reads in $\Rs$ 
will be length-$L$ sequences from the alphabet $\Sigma' = \{a,c,g,t,\vep\}$, where $\vep$ corresponds to an erasure.
Thus, 
a read starting at position $i$ from $\s$ can be written as
$\r_i = (r_i[0],...,r_i[L-1])$, where either $r_i[j] = s[i+j]$ or $r_i[j] = \vep$, for $1 \leq i \leq G$ and $0 \leq j \leq L-1$.

For a fixed parameter $D$, the adversarial erasure model will be constrained by a maximum error rate of $D/L$ within each read, and for each base.
Since in our read model each base $s[i]$ is read $L$ times ($r_{i-(L-1)}[L-1],r_{i-(L-2)}[L-2],...,r_{i}[0]$), these constraints can be written as follows:

\begin{enumerate}[ a)]
\item There are at most $D$ erasures per read.
\item Each base $s[i]$ is erased at most $D$ times across all reads.
\end{enumerate}
We will use $\Rsd(\s)$ to refer to the $L$-spectrum of $\s$, $\Rs_{L,0}(\s)$, after being corrupted by erasures satisfying (a) and (b).

In the context of an adversarial noise model with deterministic constraints, 
it makes sense to restrict our attention to
potential sequences $\hat \s$ that are \emph{consistent} with 
the reads $\Rsd(\s)$.
A sequence $\hat \s$ is said to be consistent with $\Rsd(\s)$ if it could have generated the set of reads $\Rsd(\s)$ according to the erasure model in (a) and (b).
By extension, we will say that an assembly $\sigma$ of $\Rsd(\s)$ is consistent if there exists a sequence $\hat \s$, consistent with $\Rsd(\s)$, 
that could have generated the reads in $\Rsd(\s)$ according to the positions determined by $\sigma$.
\begin{figure}[ht] 
	\center
       \includegraphics[width=0.6\linewidth]{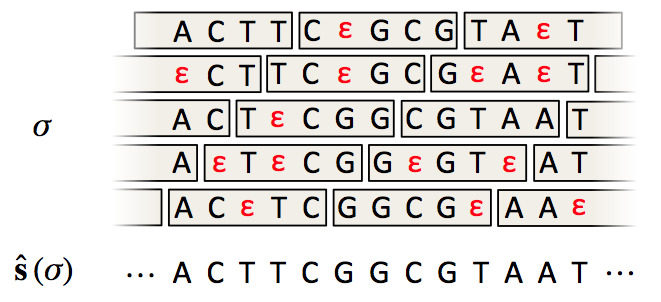}
        \caption{Part of a consistent assembly for $L = 5$ and $D = 2$. Notice that there can be at most $D$ erasures per read and per ``column'' of the assembly $\sigma$.
        Moreover, all non-erased bases in a column must agree.
        \label{consistentfig}}
        \vspace{-4mm}
\end{figure}
%
As illustrated in Fig.~\ref{consistentfig}, we notice that (b) guarantees that a consistent assembly $\sigma$ defines, up to cyclic shifts, a unique consistent sequence in $\Sigma^G$, which we will refer to as $\hat \s(\sigma)$.

The fundamental feasibility question corresponds to asking which values of $L$ allow unambiguous reconstruction.
Formally, 
it corresponds to the following algorithm-independent question.


\begin{quest}{Consider a fixed circular sequence $\s \in \Sigma^G$. 
What values of $L$ guarantee that, for an arbitrary set of erased reads $\Rsd(\s)$, $\s$ is the unique sequence consistent with $\Rsd(\s)$?}
\end{quest}

\section{Assembly in the Noiseless Case}

The assembly problem in Question 1 was first studied in \cite{Ukkonen} in the noiseless setting $D = 0$.
Notice that when $L=1$, $\Rs_{1,0}(\s)$ is simply the multi-set $\{s[1],...,s[G]\}$ and any permutation $\sigma$ of $(1,...,G)$ is a consistent assembly. 
Hence, $\s$ cannot be reconstructed unambiguously, unless all of its symbols are the same.
On the other hand, when $L=G$, there is a unique assembly of $\Rs_{G,0}(\s) = \{\s\}$, and $\s$ can always be reconstructed unambiguously.
Question 1 is thus equivalent to asking for the threshold $\lth$ for which $\s$ can be reconstructed if and only if $L > \lth$.
In \cite{Ukkonen}, this threshold is established as a function of the \emph{repeat} structure of the sequence $\s$, as we explain next.

A \emph{repeat} of length $\ell$ in $\s$ is a subsequence appearing twice at some positions $t_1$ and $t_2$ (so $\s_{t_1}^\ell$ and $\s_{t_2}^\ell$) that is maximal; i.e., $s[t_1-1] \ne s[t_2-1]$ and $s[t_1+\ell] \ne s[t_2+\ell]$.
Two pairs of repeats $\s_{a_1}^{\ell}, \s_{a_2}^{\ell}$ and $\s_{b_1}^{k}, \s_{b_2}^{k}$ are \emph{interleaved} if 
$a_1 < b_1 \leq a_2 < b_2$.
Due to the circular DNA model, since a subsequence $\s_t^\ell$ can also be written as $\s_{t+mG}^\ell$ for any integer $m$, we additionally require that $b_2 - a_1 < G$.
The length of a pair of interleaved repeats $\s_{a_1}^{\ell}, \s_{a_2}^{\ell}$ and $\s_{b_1}^{k}, \s_{b_2}^{k}$ is defined to be $\min(\ell,k)$.
We let $\linter(\s)$ be the length of the longest pair of interleaved repeats in $\s$ and set $\lcrit(\s) = \linter(\s)+1$.
The results from \cite{Ukkonen,PevznerDNA} imply the following:

\begin{theorem} \label{ukkonenthm}
If $L > \lcrit(\s)$, 
then $\s$ is the unique sequence that is consistent with $\Rs_{L,0}(\s)$.
Conversely, if $L \leq \lcrit(\s)$, 
there exists a sequence $\s' \ne \s$ that is also consistent with $\Rs_{L,0}(\s)$.
\end{theorem}

In other words, Theorem~\ref{ukkonenthm} characterizes the threshold on $L$ that fully answers Question 1.
We point out that, in the previous literature \cite{Ukkonen,BBT}, $\lcrit$ was defined in terms of the length of pairs of interleaved repeats (defined in a more restrictive way) and the length of \emph{triple repeats}.
However, one can verify that by considering the more general definition of interleaved repeats above, triple repeats are included as a special case.

Notice that, while Theorem~\ref{ukkonenthm} characterizes the minimum $L$ that guarantees perfect reconstruction, $\lcrit(\s)$ is a function of the ground truth $\s$, and is not known a priori. 
However, the following corollary of Theorem~\ref{ukkonenthm} readily follows:
\begin{cor} \label{certcor1}
If a sequence $\hat \s$ is consistent with $\Rs_{L,0}(\s)$ and $L > \lcrit(\hat \s)$, then $\hat \s = \s$.
\end{cor}
Since $\lcrit(\hat \s)$ can be computed from the assembled sequence $\hat \s$, this result means that $L > \lcrit(\hat \s)$ provides a certificate that $\hat \s = \s$, even without previous knowledge of $\lcrit(\s)$.


%

\section{Main Results}
\label{mainsec}

In the previous section, we described how Theorem~\ref{ukkonenthm} fully characterizes when assembly is possible given the noiseless $L$-spectrum.
In this section, we seek a similar characterization 
in the case where reads are noisy.

Notice that for the erasure setting described in Section~\ref{probsec}, one possible erasure pattern is to have the last $D$ bases from each read erased, which effectively results in noiseless reads of length $L-D$.
Therefore,  
the converse part of Theorem~\ref{ukkonenthm} implies that, if $L \leq \lcrit(\s) + D$, 
there is a read set $\Rsd(\s)$ 
and a sequence $\hat \s \ne \s$ that is consistent with $\Rsd(\s)$.
But how much larger than $\lcrit(\s)+D$ does the read length $L$ have to be in order to guarantee unambiguous correct reconstruction?
In other words, how do erasures degrade the fundamental limit characterized by Theorem~\ref{ukkonenthm}?

Our main result is the introduction of a new sequence-dependent quantity, $\lacrit (D, \s)$, such that, if $L > \lacrit(\s,D)$, 
$\s$ is the unique sequence consistent with $\Rsd(\s)$.
In general, $\lcrit(\s) + D < \lacrit(\s,D)$ for $D > 0$, and one can construct an arbitrary sequence $\s \in \Sigma^G$ for which the gap between the two quantities is significant.
However, by computing $\lcrit + D$ and $\lacrit$ for actual genomes, we verify that they are often close, as shown in Table~\ref{table1}. 

Rather than being defined in terms of exact repeats, as is the case of $\lcrit(\s)$, $\lacrit(\s)$ depends more generally on \emph{approximate repeats}.
For a set of segments $\S$ of a given length $\ell$; i.e., $\S \subset \Sigma^\ell$, we will first define the radius of $\S$ to be
\al{ \label{radiusdef}
\rho(\S) = \min_{\x \in \Sigma^\ell} \max_{\y \in \S} d_H (\y,\x),
}
where $d_H (\y,\x)$ is the Hamming distance between $\y$ and $\x$.
We will say that the segments in $\S$ are $d$-approximate copies if $\rho(\S) \leq d$.
Intuitively, a sequence $\s$ that contains a large set $\S$ of length-$\ell$ segments with a small radius $\rho(\S)$ 
has more ambiguity in terms of assembly.
To capture that, we will let $M(d,\ell)$ correspond to the maximum number of $d$-approximate length-$\ell$ segments in $\s$; i.e.,
\al{
M_{\s}(d,\ell) = \max \left\{ |\S| : \S \subset \Rs_{\ell,0}(\s), \rho(\S) \leq d\right\}.
}
%
%
%
%
Notice that $M_\s(d,\ell)$ is monotonically decreasing in $\ell$.
We let 
\al{ \label{lacriteq}
\lacrit(\s,D) = \min_{k \geq \lcrit(\s)} k + D \cdot M_{\s}(D,k+1).
}
Notice that $\lacrit(\s,D) \geq \lacrit(\s,0) = \lcrit(\s)$.
Our main result is the following.
\begin{theorem}\label{lacritthm}
If $L > \lacrit(\s,D)$, 
then $\s$ is the unique sequence that is consistent with $\Rsd(\s)$.
\end{theorem}

%

The main tool used to prove Theorem~\ref{lacritthm} is a result about spectrum error correction.
More precisely, we show that from a noisy version of the $L$-spectrum of $\s$ $\Rsd(\s)$, it is possible to obtain $\Rs_{L',0}(\s)$, for some effective read length $L' < L$.
This result and the proof of Theorem~\ref{lacritthm} are presented in Section~\ref{proofsec}.

As in the noiseless case, we point out that $\lacrit(\s,D)$ cannot be computed a priori, since it is a function of the ground truth sequence $\s$.
However, Theorem \ref{lacritthm} can in fact be used to obtain a certificate result analogous to Corollary~\ref{certcor1}, allowing one to certify whether an assembly $\hat{s}$ is correct, even without prior knowledge of $\lacrit(\s)$ and $M_{\s}(D,  \cdot)$.


\begin{cor} \label{certcor2}
If a sequence $\hat \s$ is consistent with $\Rs_{L,D}(\s)$ and $L > \lacrit(\hat \s)$, then $\hat \s = \s$.
\end{cor}

\begin{proof}
If $\hat \s$ is consistent with $\Rsd(\s)$, by the definition of consistency, $\Rsd(\s)$ can be viewed as a set of reads $\Rsd(\hat \s)$ from $\hat \s$, with an erasure pattern satisfying (a) and (b).
But from Theorem~\ref{lacritthm}, if $L > \lacrit(\hat \s)$, $\hat \s$ is the unique sequence that is consistent with $\Rsd(\hat \s) = \Rsd(\s)$.
Since $\s$ must also be consistent with $\Rsd(\hat \s)$, we must have $\hat \s = \s$.
\end{proof}

In Table~\ref{table1}, we show the value of $\lacrit(\s,D)$ computed for several real genomes.
Computing $\lacrit(\s,D)$ is generally impractical from a computational standpoint,
so the values in Table~\ref{table1} are based on heuristics implemented by a sequence alignment tool called Nucmer \cite{nucmer}.
We choose the value of $D$ such that $D/\lcrit \approx 15\%$.
We point out that the first two genomes, \emph{R. sphaeroides} and \emph{S. aureus} are from the GAGE dataset \cite{GAGE}, which is used as a benchmark for assemblers.
Notice that, with the exception of \emph{E. coli 536}, in all cases $\lacrit(\s,D) = \lcrit(\s) + m D$, for  $m \in \{2,3,4\}$.
This occurs because, for the genomes considered, $\lcrit(\s)$ is already long enough so that there aren't many approximate repeats of that length.

\begin{table}[h]
\begin{center}
\begin{tabular}{lccc}
  \toprule
  Genome ($\s$) & $\lcrit(\s)$ & $\lacrit(\s,D)$ & $D$ \\
  \midrule
  \emph{R. sphaeroides} & 271 & 331 & 30 \\
  \emph{S. aureus} & 1799 & 2399 & 200 \\
  \emph{A. ferrooxidans \hspace{1cm}} & 2628 & 3228 & 300 \\
  \emph{E. coli 536} & 3245 & 4462 & 450 \\
  \emph{E. coli K-12} & 1744 & 2544 & 200 \\
  \bottomrule
\end{tabular}
 \vspace{2mm}
  \caption{Computed $\lacrit(\s,D)$ for $D/\lcrit \approx 15\%$}
 \label{table1}
\end{center}
\vspace{-5mm}
\end{table}

\vspace{-2mm}

While the results in this section were presented for an erasure model, they can be extended to a substitution error model.
In fact, if 
instead of $D$ erasures per read and per base, we have $D/2$ substitution errors,
the proofs of Theorems~\ref{lacritthm} and \ref{ecthm} can be modified accordingly, and the statements still hold.
We will restrict the discussion to the erasure case for simplicity.

\section{Spectrum Error Correction} 
\label{proofsec}

The main result we use to prove Theorem~\ref{lacritthm} is a statement about when it is possible to take a noisy $L$-spectrum of $\s$ and unambiguously construct its noiseless $L'$-spectrum, for $L' < L$.

\begin{theorem} \label{ecthm}
Suppose that, for some $k$, we have 
\al{ \label{Leq}
L > k + D \cdot M_{\s}(D,k+1).
}
Then, for any sequence $\hat \s$ that is consistent with $\Rsd(\s)$, $\Rs_{k+1,0}(\hat \s) = \Rs_{k+1,0}(\s)$.
\end{theorem}

Theorem~\ref{ecthm} says that, by finding a consistent assembly of $\Rsd(\s)$, we can obtain the (noiseless) $(k+1)$-spectrum of $\s$, as long as $k$ satisfies \eref{Leq}.
Therefore, when $L > \lacrit(\s,D)$, if we let $k^{\star}$ be the minimizer in \eref{lacriteq}, we have that $L > k^{\star} + D \cdot M_{\s}(D,k^{\star}+1)$ and, by Theorem~\ref{ecthm}, any $\hat \s$ that is consistent with $\Rsd(\s)$ has the same $(k^{\star}+1)$-spectrum $\Rs_{k^{\star}+1,0}(\s)$.
But since, $k^{\star} + 1 > \lcrit(\s)$, Theorem~\ref{ukkonenthm} implies that there is only one sequence that is consistent with $\Rs_{k^{\star}+1,0}(\s)$, and we must have $\hat \s = \s$.
This proves Theorem~\ref{lacritthm}.

%
Next, we turn to the proof of Theorem~\ref{ecthm}.
Suppose that we pick some $k$ satisfying \eref{Leq} and
that $\sigma$ is a consistent assembly for the set of reads $\Rsd(\s)$
with assembled sequence $\hat \s = \hat \s(\sigma)$.
The main idea of the proof is to show that $(k+1)$-blocks in $\s$ and $\hat \s$ are in one-to-one correspondence; i.e., $\hat \s_t^{k+1} = \s_{\tau(t)}^{k+1}$ for a bijective mapping $\tau : \{1,...,G\} \to \{1,...,G\}$, which implies
$\Rs_{k+1,0}(\s) = \Rs_{k+1,0}(\hat \s)$.

\begin{figure}[t] 
	\center
       \includegraphics[width=\linewidth]{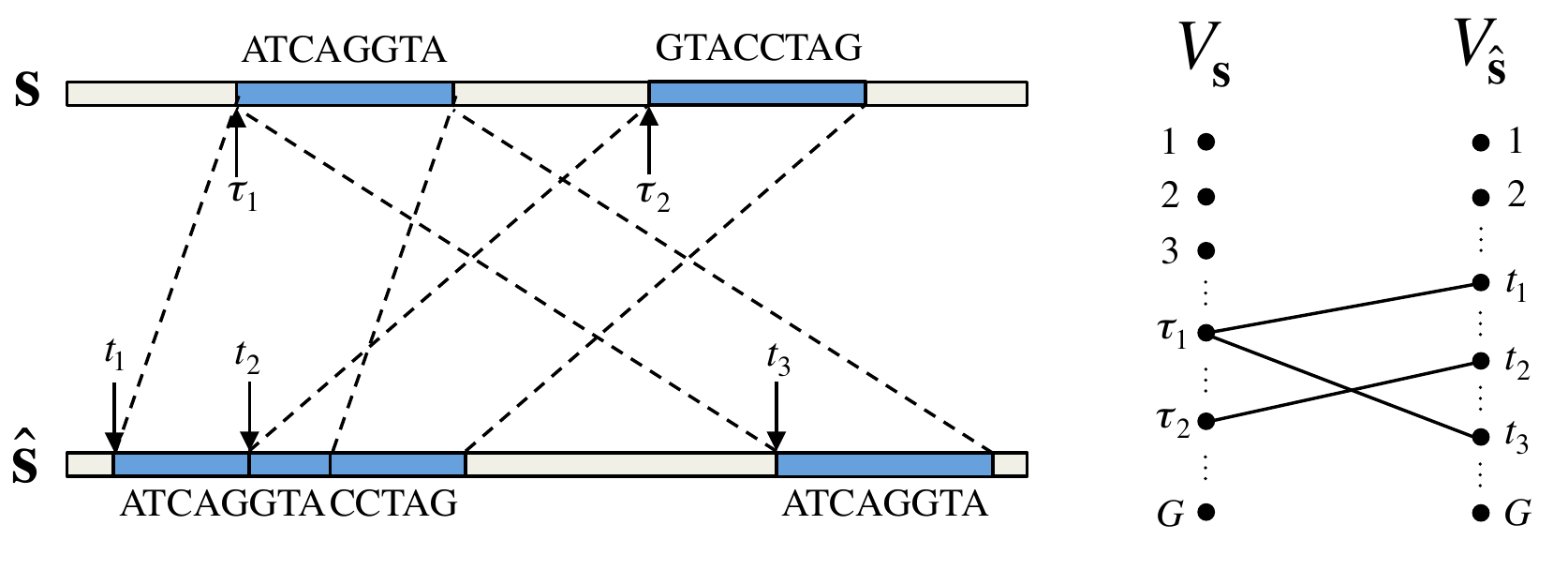}
        \caption{We place an edge $(u,v)$ in $(V_{\s},V_{\hat \s},E)$ if $\s_{u}^{k+1} = \hat \s_{v}^{k+1}$.
        In this example, $(\tau_1,t_1)$, $(\tau_2,t_2)$ and $(\tau_1,t_3)$ are some of the edges in $(V_{\s},V_{\hat \s},E)$.        
        \label{graphfig}}
        \vspace{-4mm}
\end{figure}

In order to show the existence of this bijection $\tau$,
we consider 
a bipartite graph $(V_{\s},V_{\hat \s},E_{k+1})$, where 
$V_{\s} = V_{\hat \s} = \{1,...,G\}$ 
and 
$E = \{ (u,v) \in V_{\s} \times V_{\hat \s} : \s_{u}^{k+1} = \hat \s_{v}^{k+1} \}$,
as illustrated in Fig.~\ref{graphfig}.
The existence of the bijective mapping $\tau$ is equivalent to the existence of a perfect matching in $(V_{\s},V_{\hat \s},E)$.
Hence, Theorem~\ref{ecthm} is equivalent to the following:


\begin{claim} \label{pmclaim}
There exists a perfect matching in $(V_{\s},V_{\hat \s},E)$.
\end{claim}

For a set of nodes $U \subset V_{\hat \s}$, we let $\delta(U) = \{v \in V_{\s} : (v,u) \in E \text{ for $u \in U$} \}$ be the set of neighbors of  $U$.
We will show that, for any $U \subset V_{\hat \s}$, $|\delta(U)| \geq |U|$, and by Hall's marriage theorem, Claim~\ref{pmclaim} will follow. 
We will first state the following lemma, which establishes $|\delta(U)| \geq |U|$ for the special case of sets
$U$ of the form $U_{\x} = \{ u \in V_{\hat \s} : \hat \s_u^{k+1} = \x \}$ for some $\x \in \Sigma^{k+1}$.

\begin{lemma} \label{uxlem}
For the bipartite graph $(V_\s,V_{\hat \s},E)$, 
$|\delta(U_\x)| \geq |U_\x|$, for any $\x \in \Sigma^{k+1}$.
\end{lemma}

The proof of Lemma~\ref{uxlem} is at the end of this section. 
Now consider a general set $U \in V_{\hat \s}$.
Let $\S_U^{k+1} = \{ \s_u^{k+1} \in \Sigma^{k+1} : u \in U\}$.
Since two nodes $u,u'\in U$ with $\s_u^{k+1}\ne \s_{u'}^{k+1}$ cannot be connected to the same node $v \in V_{\s}$, we have
\aln{
|\delta(U)| & = \sum_{\x \in \S_U^{k+1}} |\delta( U_{\x} \cap U) | = \sum_{\x \in \S_U^{k+1}} |\delta( U_{\x} ) | \\
&\geq \sum_{\x \in \S_U^{k+1}} |U_{\x}|  \geq \sum_{\x \in \S_U^{k+1}} |U_{\x} \cap U| = |U|,
}
where the first inequality follows from Lemma~\ref{uxlem}.
By applying Hall's theorem, Claim~\ref{pmclaim} follows,
implying that, $\Rs_{k+1,0}(\s) = \Rs_{k+1,0}(\hat \s)$. 
Therefore, to conclude the proof of Theorem~\ref{ecthm}, we just need to prove Lemma~\ref{uxlem}.

\begin{proof}[Proof of Lemma~\ref{uxlem}]
Let $U_{\x} = \{t_1,...,t_q\} \subset V_{\hat \s}$, where 
$t_1,...,t_q$ are distinct and
$\hat \s_{t_1}^{k+1} = ... = \hat \s_{t_q}^{k+1} = \x$.
Consider one such block $\hat \s_{t}^{k+1}$, for $t \in \{t_1,...,t_q\}$.
There are $L-k$ reads that cover $\hat \s_{t}^{k+1}$ in $\hat \s$, as illustrated in Fig.~\ref{proof1fig}.
These are the reads given by  
$\r_{\sigma^{-1}(t-n)}$, for $n=0,1,...,L-k-1$.
\begin{figure}[t] 
	\center
       \includegraphics[width=0.9\linewidth]{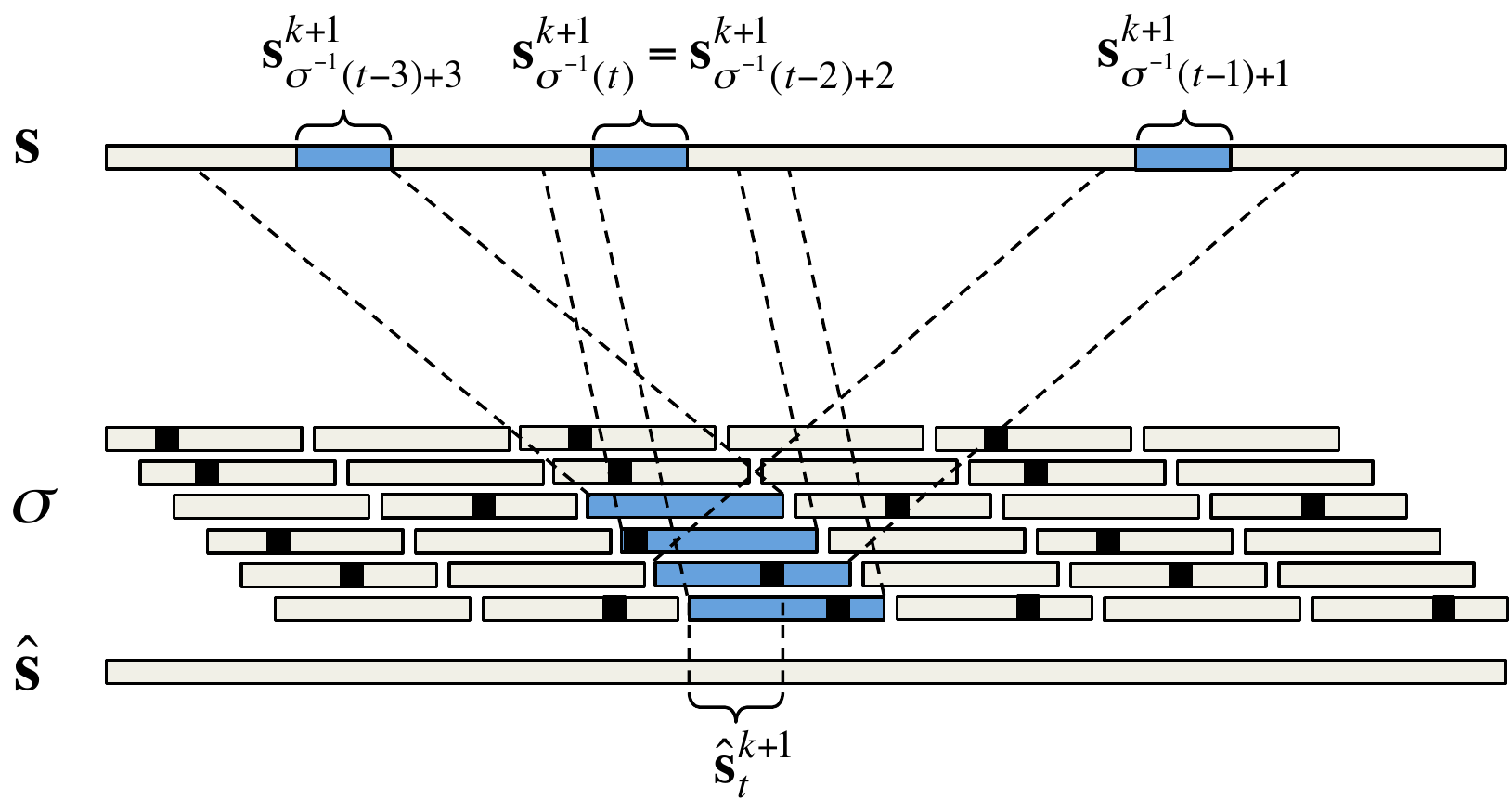}
       \vspace{-2mm}
        \caption{For an arbitrary length-$(k+1)$ block $B$ in $\hat s$, the $L-k$ reads that completely cover $B$ according to the assembly $\sigma$ are shaded (in this example, $L = 6$ and $k = 2$).
By mapping these $L-k$ reads back to $\s$, we find the corresponding $(k+1)$-blocks in $\s$ given by $\s_{\sigma^{-1}(t-j)+j}^{k+1}$ for $j=0,1,2,3$.
Notice that, in this example, $\sigma^{-1}(t) = \sigma^{-1}(t-2)+2$, 
because reads $\r_{\sigma^{-1}(t)}$ and $\r_{\sigma^{-1}(t-2)+2}$ are aligned to each other in the same way in $\s$ and $\hat \s$.
        \label{proof1fig}}
        \vspace{-4mm}
\end{figure}
Notice that read $\r_{\sigma^{-1}(t-n)}$ was originally obtained from the segment $\s_{\sigma^{-1}(t-n)}^L$ from the true sequence $\s$.
The consistency requirement on $\sigma$ thus implies that $d_H( \s_{\sigma^{-1}(t-n)}^L, \s_{t-n}^L) \leq D$.
Moreover, if we just focus on the $(k+1)$-block corresponding to $\hat \s_t^{k+1}$, we have
$d_H( \s_{\sigma^{-1}(t-n)+n}^{k+1}, \s_{t}^{k+1}) = d_H( \s_{\sigma^{-1}(t-n)+n}^{k+1}, \x ) \leq D$, which holds for each $t \in \{t_1,...,t_q\}$ and $n=0,...,L-k-1$.


If we now consider the set of all such $(k+1)$-blocks in $\s$ 
\al{ \label{setS2}
\S = \left\{ \s_{\sigma^{-1}(t_i-n)+n}^{k+1} : n = 0,...,L-k-1, i=1,...,q \right\},
}
since $d_H( \y, \x ) \leq D$ for each $\y \in \S$, we have that $\rho(\S) \leq D$.
%
%
%
%
%
Hence, if we let 
\aln{
\T = \left\{ \sigma^{-1}(t_i-n)+n : 0 \leq n \leq L-k-1, 1 \leq i \leq q  \right\} 
}
be the starting positions of these blocks in $\s$, $\T$
must satisfy $|\T| \leq M_{\s}(D,k+1)$.
Now consider the set of $(n,i)$ pairs
\aln{
\B = \{ (n,i) : 0 \leq n \leq L-k-1, 1 \leq i \leq m\}.
}
We will define a partition on $\B$ according to the value of $\sigma^{-1}(t_i-n)+n$.
More precisely, we will let
%
%
\aln{
\B_{\tau} = \left\{ (n,i) \in \B : \sigma^{-1}(t_i-n)+n = \tau  \right\},
}
for $\tau \in \T$.
It is clear that $\{ \B_\tau \}_{\tau \in \T}$ is a partition of $\B$.
We claim that there exist distinct $\tau_1,...,\tau_q \in \T$ such that 
$|\B_{\tau_j} | \geq D+1$, for $j=1,...,q$.
Suppose by contradiction that this is not the case, and we have at most $q-1$ parts $\B_\tau$ with $|\B_\tau| \geq D+1$.
Notice that, since $\sigma : (1,...,G) \to (1,...,G)$ is one-to-one, $\sigma^{-1}(t_i-n)+n \ne \sigma^{-1}(t_j-n)+n$ if $t_i \ne t_j$, and, for any $\tau$, we must have $|\B_\tau| \leq L - k$. 
Therefore, since \eref{Leq} implies $L-k - 1\geq D\cdot M_{\s}(D,k+1)$,
\aln{
\sum_{\tau \in \T} |\B_\tau| & \leq (q-1) (L-k) + (|\T|-q+1) D \\ 
& \leq (q-1) (L-k) + D \cdot M_{\s}(D,k+1) \\
& = q (L-k) - 1.
}
But since $\sum_{\tau \in \T} |\B_\tau| = |\B| = q (L-k)$, we have a contradiction.

Now consider the segments $\s_{\tau_j}^{k+1}$ with $|\B_{\tau_j}|\geq D+1$, for $j=1,...,q$.
Since  $\tau_1,...,\tau_q$ are all distinct, these segments  start at different points in $\s$.
Moreover, since $|\B_{\tau_j}|\geq D+1$, each $\s_{\tau_j}^{k+1}$ is covered by $D+1$ reads from the reads that cover $\hat \s_{t_i}^{k+1}$, $i=1,...q$. 
Notice that these must be distinct reads from the multiset $\Rsd(\s)$.
This is because two distinct pairs $(n,i)$ and $(m,j)$ in $\B_{\tau}$ must have $n \ne m$, and the corresponding reads are $\r_{\sigma^{-1}(t_i-n)} = \r_{\tau - n}$ and $\r_{\sigma^{-1}(t_{j}-m)} = \r_{\tau - m}$, which are distinct reads (not necessarily different sequences from $\Sigma^L$).
Finally, as illustrated in Fig.~\ref{proof3fig}, we note that, since there are at most $D$ erasures per base in $\s$, we have that $\s_{\tau_j}^{k+1} = \x$, for $j=1,...,q$. 
We conclude that $|\delta(U)| \geq q$.
\begin{figure}[t] 
	\center
       \includegraphics[width=0.74\linewidth]{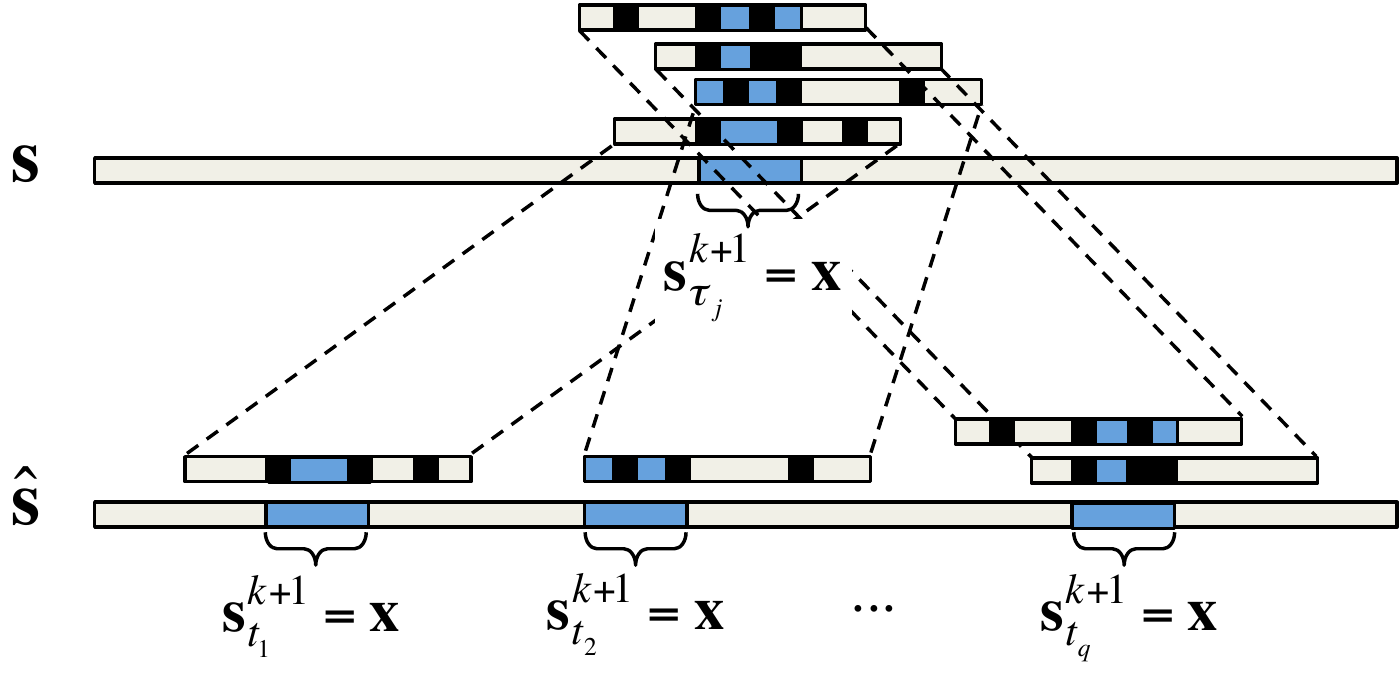}
        \caption{If $|\B_{\tau_j}| \geq D+1$, at least $D+1$ of the reads that cover one of $\hat \s_{t_1}^{k+1},...,\hat \s_{t_q}^{k+1}$ in $\hat \s$ also cover $\s_{\tau_j}^{k+1}$ in $\s$ (in this example, $D = 3$).
        Since there are at most $D$ erasures per base in $\s$, we must have $\s_{\tau_j}^{k+1} = \x$.
        \label{proof3fig}}
        \vspace{-4mm}
\end{figure}
\end{proof}



\section{Concluding Remarks}

Our results show that for several actual genomes, if we are in a dense-read model with reads $20$-$40 \%$ longer than the noiseless requirement $\lcrit(\s)$, perfect assembly feasibility is robust to erasures at a rate of about $10\%$.
While this is not as optimistic as the message from \cite{MotahariNoisy}, we emphasize that we consider an adversarial error model.
When errors instead occur at random locations, it is natural to expect less stringent requirements.

Another message provided by our results deals with error correction.
Most current sequencing technologies employ error correction algorithms based on aligning reads to form clusters and outputing a cleaned-up read for each cluster.
However, the spectrum error correction result from Theorem~\ref{ecthm} suggests that a ``global'' approach to generating cleaned-up reads (based on finding a consistent assembly and looking at its spectrum) may perform better than cluster-based, or local, error correction.

A direction for future work is to replace the dense-read model with a shotgun read model.
While the $L$-spectrum approach is motivated by the high-throughput of current technologies, 
it bypasses the question of the actual coverage depth required for assembly.
As was the case in \cite{BBT}, we expect the read length requirements from the dense-read model to translate into \emph{bridging} conditions in the shotgun model, allowing one to compute the coverage required for perfect reconstruction with high probability.

%
%
%
%
%
%


%

\section*{Acknowledgment}
This work is partially supported by the Center for Science of Information (CSoI), an NSF
Science and Technology Center, under grant agreement CCF-0939370.



%
%
%

{\footnotesize
\bibliographystyle{IEEEtran}
\bibliography{../../../refsc}
}

\end{document}